\documentclass[aps,prb,superscriptaddress,reprint,floatfix,longbibliography,amsfonts,amsmath,amssymb]{revtex4-2}
\usepackage{graphicx}
\usepackage{hyperref}
\usepackage{epstopdf}
\usepackage[note-name=, use-sort-key = false]{notes2bib}
\usepackage{color}
\usepackage{pifont}
\usepackage{pdfpages}
\usepackage[version=4]{mhchem}

\hypersetup{
    colorlinks = true,
    citecolor = blue,
    bookmarksnumbered = true,
    urlcolor = blue
}

\makeatletter
\AtBeginDocument{\let\LS@rot\@undefined}
\makeatother

\newcommand{\mn}{\ce{Mn3Si2Te6}}
\newcommand{\cgt}{\ce{CrGeTe3}}
\newcommand{\tc}{$T_{\rm C}$}

\begin{document}

\title{Origin of pressure-induced anomalies in the nodal-line ferrimagnet {\mn}}

\author{Varun Venkatasubramanian}
\affiliation{Research Institute for Interdisciplinary Science, Okayama University, Okayama 700-8530, Japan}

\author{Makoto Shimizu}
\affiliation{Department of Physics, Graduate School of Science, Kyoto University, Kyoto 606-8502, Japan}

\author{Daniel Guterding}
\email{daniel.guterding@th-brandenburg.de}
\affiliation{Technische Hochschule Brandenburg, Magdeburger Straße 50, 14770 Brandenburg an der Havel, Germany}

\author{Harald O. Jeschke}
\email{jeschke@okayama-u.ac.jp}
\affiliation{Research Institute for Interdisciplinary Science, Okayama University, Okayama 700-8530, Japan}

\begin{abstract}
A pressure-induced insulator-to-metal transition (IMT) has recently been discovered in the nodal-line ferrimagnet {\mn}. The electronic phase transition is accompanied by anomalies in the magnetic ordering temperature and the anomalous Hall conductivity, which peak at or near the critical pressure of the IMT. We perform density functional theory (DFT) calculations as a function of pressure to establish the connection between the IMT and the magnetic anomalies in {\mn}. We extract Heisenberg Hamiltonians as a function of pressure based on our DFT calculations. Our classical Monte Carlo simulations for these Hamiltonians yield ordering temperatures and magnetic ordering patterns, in agreement with the experimental data. Although we can accurately explain the evolution of magnetism with pressure, it seems that the anomalous Hall conductivity in {\mn} can only be accounted for by extrinsic contributions or moderate electron doping of the samples in the experiment.
\end{abstract}

\maketitle

\section{Introduction}
Magnetic materials with non-trivial band topology have recently emerged as a platform for novel transport phenomena with potential spintronic applications~\cite{Manna2018, Tokura2019, Nagaosa2020, Nakatsuji2022, Bernevig2022}. In particular, nodal-line semiconductors and semimetals exhibit a particularly large anomalous Hall effect~\cite{Liu2018b, Wang2018, Kim2018} and angular magnetoresistance~\cite{Ali2016, Suzuki2019, Hassinger2019, Seo2021, Das2025}. 

Among them, {\mn} stands out as a ferrimagnetic nodal-line semiconductor that exhibits record-high colossal magnetoresistance (CMR)~\cite{Ni2021} and angular magnetoresistance (AMR)~\cite{Seo2021}. These phenomena have been linked to chiral orbital currents~\cite{Zhang2022} and highlight the material as a candidate system where magnetism, topology, and electronic correlations are strongly intertwined.

At ambient pressure, {\mn} crystallizes in a trigonal structure (space group P$\overline{3}1$c)~\cite{Vincent1986} and orders ferrimagnetically below $T_{\rm C} \approx 78$~K~\cite{May2017, Liu2018, May2020}. Previous studies suggest that its magnetic and electronic properties are closely coupled, with signatures of a field-driven insulator-to-metal transition~\cite{Gu2024} and current-driven effects on magnetism~\cite{Zhang2024}. Single-crystal neutron diffraction shows a slight non-collinearity of magnetic moments, which points to a delicate frustration of magnetic couplings~\cite{Ye2022}.

The magnetic anisotropy at ambient pressure has been estimated as $\sim 0.7$~meV/Mn~\cite{Bigi2023, Zhang2023}, which translates to $\sim 1$~K after dividing out the spin factor $S^2 = 25/4$. The spin waves of {\mn} at ambient pressure have been mapped with inelastic neutron scattering. A fit of the spectrum with isotropic interactions plus single-ion anisotropy also yields $\sim 1$~K~\cite{Sala2022}. One study also finds a Dzyaloshinskii–Moriya interaction~\cite{Bigi2023}.

Application of pressure reveals even richer physics. At $P_c=15.4$\,GPa, {\mn} undergoes a structural transition to a monoclinic phase, concurrent with an insulator-to-metal transition~\cite{Susilo2024}. A recent study places the critical pressure at a slightly lower pressure of 10--12~GPa~\cite{Hou2025}. From ambient to critical pressure, the ferrimagnetic ordering temperature increases almost linearly to nearly room temperature, before decreasing again at higher pressures, resulting in a dome-shaped evolution of {\tc}~\cite{Susilo2024}. After the transition to the monoclinic crystal structure, a pronounced anomalous Hall effect emerges, which peaks near 17~GPa~\cite{Susilo2024}. These correlated anomalies strongly suggest that changes in the electronic structure under compression are intimately linked to the evolution of magnetism.

Here, we investigate the microscopic origin of these pressure-induced anomalies. Using density functional theory (DFT) calculations combined with classical Monte Carlo simulations, we extract exchange couplings, ordering temperatures and anisotropies across the structural transition, as well as the intrinsic anomalous Hall conductivity. Our results provide a comprehensive picture of how pressure tunes the interplay between electronic and magnetic degrees of freedom in {\mn}.

\begin{figure*}[t]
    \includegraphics[width=\textwidth]{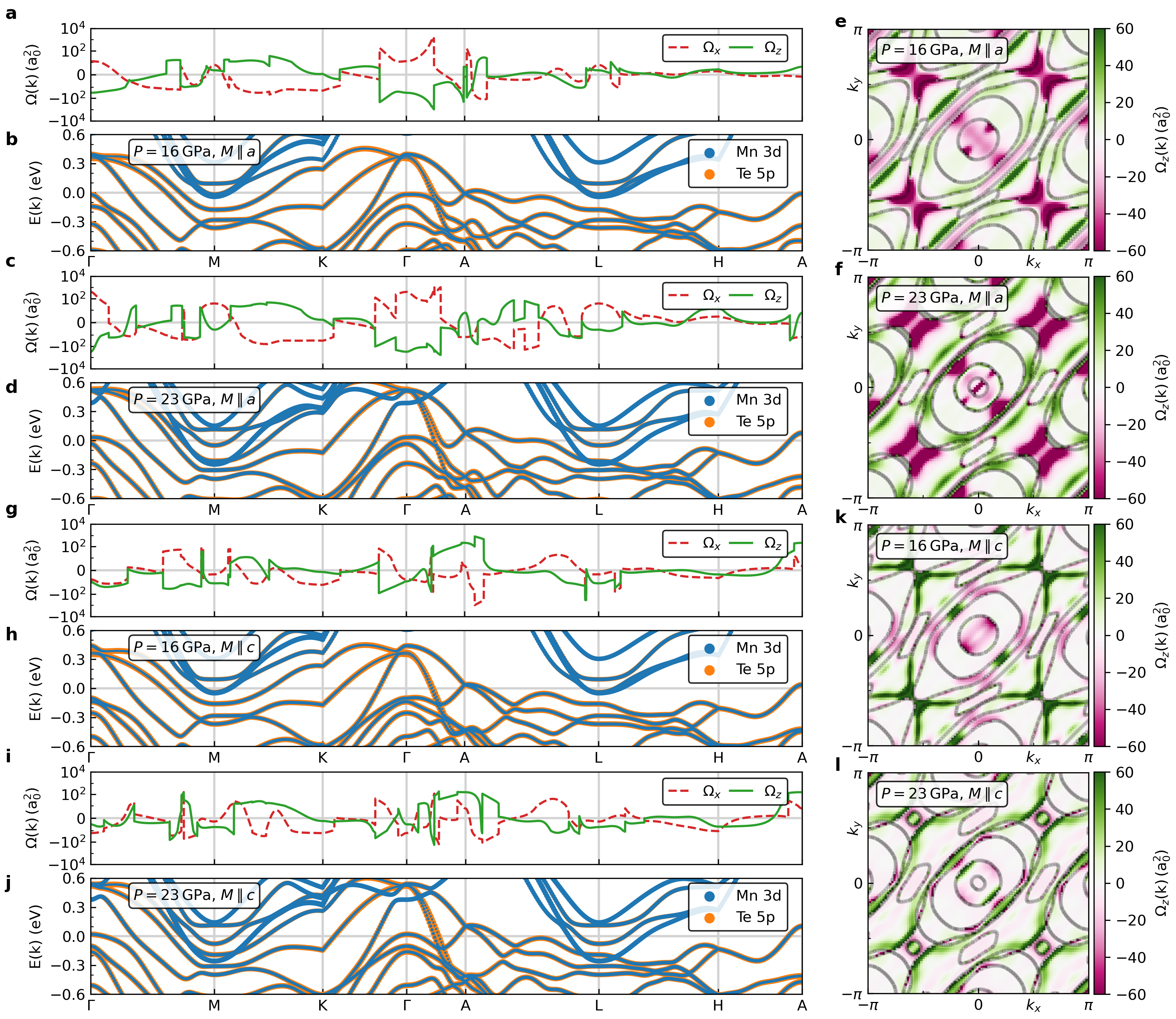}
    \caption{\textbf{Electronic structure and total Berry curvature of {\mn} in the ferrimagnetic ground state as a function of pressure and spin quantization axis.} (a,c) show the total Berry curvature $\Omega_x$ and $\Omega_z$ (in units of squared Bohr radii $a_0^2$) and (b,d) show the electronic band structure with orbital weights on a high-symmetry path through the Brillouin zone for spin quantization axis parallel to the $a$ direction at a pressure of 16 and 23~GPa, respectively. (e) and (f) show cuts of the total Berry curvature $\Omega_z$ in the $k_x$-$k_y$ plane at $k_z=0$ for the same pressures and the same orientation of spin quantization axis. The colour scale is cut off at a value of $\pm 60$. The grey shaded area represents the Fermi surface. (g-l) show the electronic structure and total Berry curvature for spin quantization axis parallel to the $c$ direction and all other parameters equal to (a-f).}
    \label{fig:ahe}
\end{figure*}

\begin{figure*}[t]
    \includegraphics[width=\textwidth]{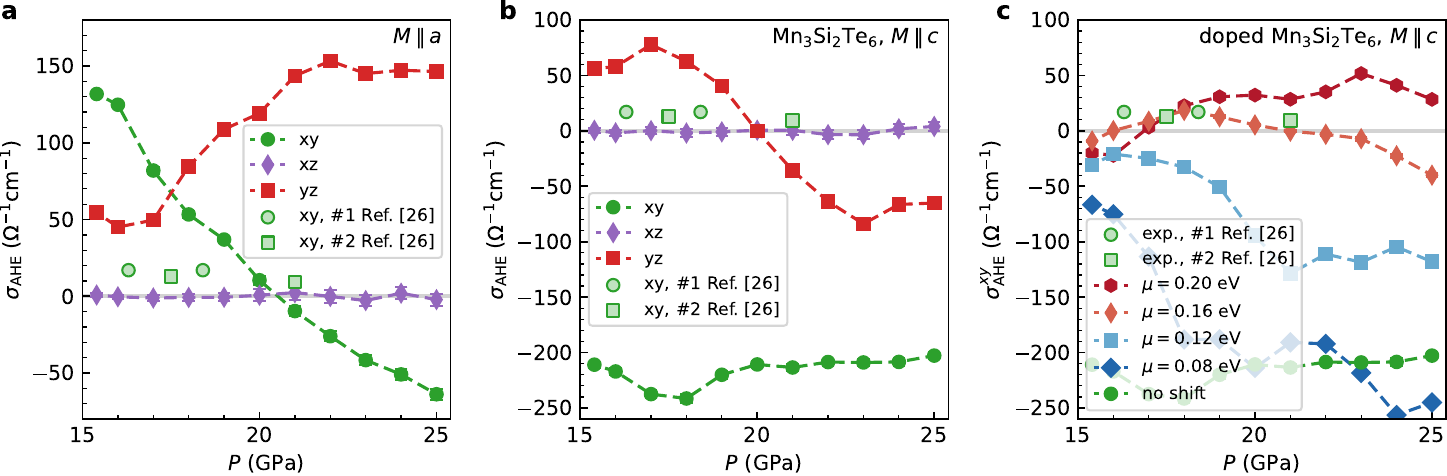}
    \caption{\textbf{Anomalous Hall conductivity of {\mn} in the ferrimagnetic ground state as a function of pressure and spin quantization axis.} (a) and (b) show the intrinsic contribution to the anomalous Hall conductivity in the $xy$, $xz$ and $yz$ planes as a function of pressure calculated from relativistic DFT calculations (i.e.~including spin-orbital coupling) in the ferrimagnetic state with spin quantization axis parallel to the $a$ direction (easy axis) and the $c$ direction (hard axis), respectively. (c) shows the calculated AHC in the $xy$ plane with shifted chemical potential $\mu$, which simulates charge doping of {\mn}, in comparison to the experimental data from Ref.~\protect\onlinecite{Susilo2024}.}
    \label{fig:ahcwithdoping}
\end{figure*}

\section{Results}
For pressures up to $P=15.4$\,GPa in the trigonal space group $P\bar{3}1c$, we determine the crystal structures by DFT structure relaxation applied to the internal coordinates in combination with experimental lattice parameters. We proceed this way because in the case of strongly correlated materials like {\mn}, structure prediction under hydrostatic pressure is not always reliable. We use a GGA+$U$ functional for a proper description of strong electronic interactions of the $3d$ electrons of the Mn$^{2+}$ ions. We fix the value of the Hund's rule coupling at $J_{\rm H}=0.76$\,eV~\cite{Mizokawa1996}. The relaxed structural parameters are not strongly dependent on the precise value of the on-site interaction $U$; for this reason, we use a generic value of $U=5$\,eV 
which is between $U=4.5$\,eV describing \ce{MnSc2S4}~\cite{Iqbal2018} and $U=5.2$\,eV that works well for \ce{Na3Mn(CO3)2Cl}~\cite{unpublished}. For the crystal structures, we use the lattice parameters determined experimentally in Ref.~\onlinecite{Susilo2024}. We interpolate the lattice parameters in order to obtain a regular and dense mesh of pressure values.

For the trigonal structures, we obtain the internal positions by DFT+$U$ structure relaxation in the ferrimagnetic state that is known to be the ground state. Note, however, that while the structures differ significantly between relaxation in non-magnetic and magnetic states, the magnetic order (ferromagnetic or ferrimagnetic) does not play a significant role. We compared the structures we obtained at $P=0$\,GPa by relaxing internal coordinates with GGA+$U=5$\,eV in ferromagnetic and ferrimagnetic states. We find that Mn-Te distances differ by less than 0.006\,{\AA}, Mn-Te-Mn angles by less than 0.01 degrees and the total energy by less than 0.4\,meV per formula unit. This indicates that the choice of magnetic order during relaxation does not have a significant influence. More details on the structures are given in the Supplementary Information~\cite{SI}.

Preparation of the monoclinic high pressure structures is more involved. We again interpolate the lattice parameters of the candidate $C2c$ structures between $P=15.4$\,GPa and 25\,GPa that were experimentally determined to be the best candidate for the high pressure space group~\cite{Susilo2024}. However, we found that DFT relaxation of internal parameters took the structure too far away from the single fully determined crystal structure at $P=22.3$\,GPa. This happens independent of the exchange correlation functional and is comparable to the situation in {\cgt}, where we similarly found that the experimental crystal structure cannot be precisely reproduced by any DFT functional~\cite{Xu2023}. For this reason, we keep the internal structure parameters constant at the values determined experimentally for $P=22.3$\,GPa.

In fact, the crystal structure of {\mn} is similar to van der Waals chalcogenides: layers of MnSiTe$_3$ are self-intercalated with Mn atoms, leading to alternating honeycomb and triangular Mn layers~\cite{Vincent1986, Susilo2024}. At the critical pressure of $P_c=15.4$\,GPa, the monoclinic phase transition splits the tellurium positions into three inequivalent sites and the MnSiTe$_3$ layers slide slightly with respect to each other~\cite{Susilo2024}.

\begin{figure*}[t]
    \includegraphics[width=\textwidth]{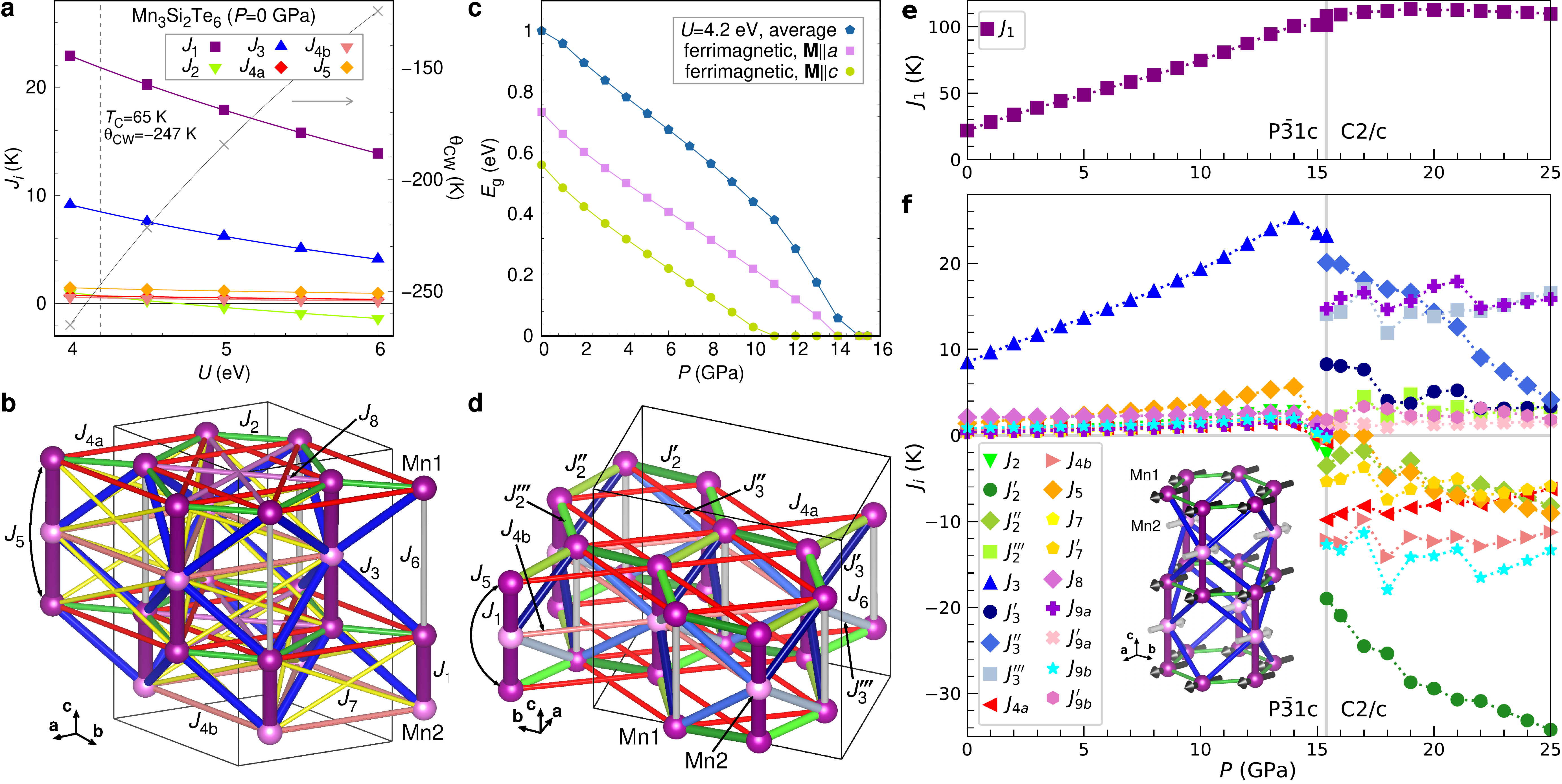}
    \caption{\textbf{DFT energy mapping and exchange pathways for {\mn}.} (a) DFT calculated exchange parameters for {\mn} at ambient pressure as a function of the on-site Coulomb repulsion $U$. The dashed line denotes the value of on-site Coulomb repulsion $U=4.2$\,eV, which we choose for the remainder of our study. For this interaction strength, we obtain $\Theta_\mathrm{CW} = -247$\,K as the mean-field Curie-Weiss temperature and $T_\mathrm{C} = 65$\,K as the ordering temperature in classical Monte Carlo, which both agree well with experimental values. (b) Relevant exchange paths in {\mn}, shown for the high symmetry, trigonal $P\bar{3}1c$ space group. (c) Charge gap of the ferrimagnetic ground state and average charge gap of all magnetic configurations included in the DFT energy mapping as a function of pressure. In the high-pressure phase, the charge gap is zero. (d) Relevant exchange paths in {\mn}, shown for the low symmetry, monoclinic $C2/c$ space group. (e) Antiferromagnetic intra-trimer (nearest neighbour) exchange coupling $J_1$ as a function of pressure. (f) Next-nearest neighbour and longer range exchange couplings between Mn atoms as a function of pressure. The interaction parameters are $U=4.2$\,eV and $J_{\rm H}=0.76$\,eV.} 
    \label{fig:couplings}
\end{figure*}

Our DFT calculations with ferrimagnetic spin configuration reproduce the experimentally observed IMT of {\mn} at a pressure of $P_c=15.4$\,GPa. For lower pressures, the system is a band insulator. For higher pressures, the system becomes metallic. The Berry curvature in the metallic phase shows a complex momentum structure with positive and negative contributions of high absolute value (see Fig.~\ref{fig:ahe}a, c, g and i). In the metallic phase, both the electronic band structure (see Fig.~\ref{fig:ahe}b, d, h, and j) and the corresponding momentum-space structures of the Berry curvature (see Fig.~\ref{fig:ahe}e, f, k, and l) evolve gradually as a function of pressure. Band structures for pressures below $15.4$~GPa are shown in Supplementary Note 1 with Supplementary Figures 4 to 6. In these we observe an insulating ferrimagnetic ground state with spins oriented in the $ab$ plane. For spins oriented along the $c$ axis, the IMT occurs below the structural phase transition, reminiscent of the field-driven IMT in {\mn}.~\cite{Gu2024} This shows how both pressure and magnetic field can be used to manipulate the electronic structure of this material.

\newcommand{\yes}{\ding{51}}
\newcommand{\no}{\ding{56}}
\begin{table*}
    \centering
    \caption{\textbf{Significant exchange interactions in {\mn}}. Properties of seven Mn-Mn bonds and how they split at the trigonal to monoclinic phase transition. $J_1$ to $J_7$ are ordered in ascending Mn-Mn distance for the ambient pressure structure.}
 \label{tab:bondidentification}
\begin{tabular}{|p{1.6cm}|p{3.0cm}|p{2.6cm}|p{2.4cm}|p{6.0cm}|}
\hline
{\bf exchange coupling} & {\bf inequivalent sites Mn$i$-Mn$j$} & {\bf $P\bar{3}m1\to C2/c$ bond splitting \mbox{$J_i\to J_i',J_i'',J_i'''$}} & {\bf bonds $J_{ia},J_{ib}$ inequivalent by symmetry} & {\bf description}\\
\hline
$J_1$ & Mn1-Mn2 & \no & \no & NN coupling forming trimers along \textit{c}\\
\hline
$J_2$ & Mn1-Mn1 & \yes & \no & \textit{ab} plane honeycomb coupling\\
\hline
$J_3$ & Mn1-Mn2 & \yes & \no & \textit{ab} plane intertrimer coupling connecting different layers\\
\hline
$J_4$ & \mbox{Mn1-Mn1}/\mbox{Mn2-Mn2} & \no & \yes & \textit{ab} plane honeycomb 2NN/triangular lattice NN\\
\hline
$J_5$ & Mn1-Mn1 & \no & \no & Mn1 2NN along $c$ (with Mn2 in between) \\
\hline
$J_6$ & Mn1-Mn1 & \no & \no & Mn1 2NN along $c$ (no Mn2 in between) \\
\hline
$J_7$ & Mn1-Mn2 & \no & \no & \textit{ab} plane intertrimer coupling forming a buckled honeycomb lattice\\
\hline
$J_8$ & Mn1-Mn1 & \no & \no & \textit{ab} plane honeycomb 3NN\\
\hline
$J_9$ & \mbox{Mn1-Mn1}/\mbox{Mn2-Mn2} & \yes & \yes & \textit{ab} plane intertrimer coupling, 2NN for $J_3$  \\
\hline
\end{tabular}
\end{table*}

For the anomalous Hall conductivity (AHC) we perform full-relativistic DFT calculations including spin-orbit coupling (SOC). The AHC arises from the Berry curvature as the integral over large positive and negative contributions, i.e.~we can expect the AHC to be very sensitive to details of the electronic structure. Although changes in the AHC as a function of pressure are gradual (see Fig.~\ref{fig:ahcwithdoping}), we observe sign changes in some components of the conductivity tensor, while others are relatively stable. The xz component is zero within numerical accuracy, regardless of the choice of quantization axis (which corresponds to the direction of an applied magnetic field in experiment). In fact, the $xz$ component is exactly zero due to the presence of antiunitary symmetries $C2^\prime (y)$ and $m^\prime (y)$. The $xy$ and $yz$ components show qualitative changes when the spin quantization axis is rotated from the easy axis (along $a$, see Fig.~\ref{fig:ahcwithdoping}a) to the hard axis (along $c$, see Fig.~\ref{fig:ahcwithdoping}b). The pressure dependence of these components results from changes in details of the electronic band structure due to the increased bandwidth and spin-orbit coupling when changing the direction of the spin quantization axis. The anisotropy of AHC in {\mn} follows from the layered crystal structure, which gives rise to an anisotropic electronic band structure, and the presence of strong SOC.

Normally, the AHC is measured in large magnetic field. Therefore, which orientation of the spin quantization axis in DFT is appropriate for comparison to experiment depends on the spin-flop field of {\mn}, which we determine together with other magnetic properties of this material. In addition, we have calculated the AHC with spins parallel to the $c$ axis under a shift of the chemical potential to simulate charge doping of the {\mn} samples. In Fig.~\ref{fig:ahcwithdoping}c we show the xy component of the AHC for moderate electron doping in comparison to theoretical values without shift and the experimental results. Electron doping strongly alters the results and a moderate doping level of $\mu=0.16$~eV yields a dome-like AHC similar to experiment. To scan the range of achievable AHC, we performed our calculations at fixed chemical potential $\mu$, which (due to pressure-induced changes in the electronic structure) is not the same as a fixed doping level. Comparison to future experimental results with a precisely known doping level will have to take this into account.

Next, we establish the Heisenberg Hamiltonian parameters for {\mn} as a function of pressure. We base our calculations on experimental crystal structures as described above. In order to determine the exchange interactions, we now apply the DFT energy mapping technique. This involves DFT calculations for a large number of spin configurations in low symmetry structures of {\mn} and fitting the total energies with the Heisenberg Hamiltonian
\begin{equation}
    H=\sum_{i<j} J_{ij}{\bf S}_i\cdot {\bf S}_j \,,
\label{eq:isohamil}
\end{equation}
where ${\bf S}_i$ are spin operators, and we do not double count bonds. This approach regularly yields excellent results in magnetic insulators~\cite{Gonzalez2024,Jaubert2025} and it has also been useful for understanding the magnetism in semiconducting and metallic {\cgt} under pressure~\cite{Xu2023, EbadAllah2025}. In fact, the latter material shares many similarities with {\mn}, for example the structural elements of the \ce{Ge2Te6} or \ce{Si2Te6} units and the observation of an anomalous Hall effect under pressure~\cite{Susilo2024,Scharf2025}. Examples for alternative approaches to magnetic Hamiltonian parameters are the four-state method~\cite{Xiang2011}, which we do not employ here because we rely on the safety net provided by our statistical approach, and the force theorem method as implemented, for example, in the \texttt{tb2j} package~\cite{He2021}; we applied the latter to ferrimagnetic {\mn} but did not arrive at a meaningful result. 

In the DFT energy mapping approach, there are two parameters that affect the overall energy scale of the exchange couplings, the on-site interaction $U$ and the Hund's rule coupling $J_{\rm H}$. The latter is expected to have little material dependence, and we fix it to $J_{\rm H}=0.76$\,eV as suggested in Ref.~\cite{Mizokawa1996}. For the on-site Coulomb interaction $U$, we exploit experimental observations at ambient pressure and then fix $U$, assuming that it is not strongly pressure dependent. The experimental 
Curie-Weiss temperature $T_{\rm CW}=-277$\,K and the ferrimagnetic ordering temperature $T_{\rm C}=78$\,K at ambient pressure~\cite{May2017} are well reproduced in our theoretical calculations if we choose $U=4.2$\,eV for this entire study (see Fig.~\ref{fig:couplings}a). With the fixed parameters $U=4.2$\,eV and $J_{\rm H}=0.76$\,eV, we now calculate the Heisenberg Hamiltonian parameters; they are shown in Fig.~\ref{fig:couplings}b, d, e and f. Table~\ref{tab:bondidentification} identifies the most important bonds and specifies how they evolve across the structural phase transition.

We find that in the trigonal low pressure region, Hamiltonians are dominated by antiferromagnetic $J_1$ and $J_3$. $J_1$ mediates the exchange between two Mn1 and one Mn2 in the nearest-neighbour Mn trimer; in this sense, the network defined by $J_1$ is zero-dimensional, and below the ordering temperature for $J_1$, the trimers order in up-down-up or down-up-down states and the effective moment of the system is reduced to one third. The existence of a large intra-trimer coupling $J_1$ is consistent with experiments that show signs of short-range magnetic order well above the ordering temperature $T_{\rm C}$~\cite{May2017, Ye2022, Baral2025}. Mn trimers form a triangular lattice in the $ab$ plane, and triangle centers define the position to which triangular lattices above and below are shifted. As a consequence, $J_3$ couples the Mn trimers into a buckled honeycomb lattice in the $ab$ plane. However, as this occurs above and below each triangular lattice of trimers, $J_3$ provides a fully three-dimensional connectivity of the lattice. It couples Mn1 and Mn2 sites, and therefore the Neel state formed by unfrustrated $J_3$ corresponds to a ferromagnetic order of the trimers with effective spin of $S_{\rm eff}=\frac{5}{2}$. The resulting ferrimagnetic state is an antiparallel arrangement of Mn1 and Mn2 sublattices which occur in a ratio 2:1 (see inset of Fig.~\ref{fig:couplings}\,f). Therefore, the nearly linear increase of $J_3$ in the insulating phase up to $P=15.4$\,GPa indicates that we can expect a similar linear increase for the ferrimagnetic ordering temperature of the material. Note that our calculations do not support a substantial $J_2$ which we find to be 3\% of $J_1$ rather than the 18\% obtained in Ref.~\onlinecite{May2017}.

We confirm this expectation by classical Monte Carlo (cMC) simulation of the full Hamiltonians shown in Fig.~\ref{fig:couplings}. In Fig.~\ref{fig:Tc} we show that the calculated {\tc} from our Hamiltonians can explain the {\tc} increase seen in experimental resistivity and susceptibility data from Ref.~\onlinecite{Susilo2024} very well (for details see Supplementary Note 2 with Supplementary Figures 7 to 12). In our calculations, the pressure induced insulator to metal transition in the ferrimagnetic ground state (with spins oriented along the $a$ axis) occurs between 14 and 16\,GPa (see Fig.~\ref{fig:couplings}c). With spins quantized along the $c$ axis (hard axis), the IMT occurs somewhat earlier, indicating the field-tunability of this material. Between 14 and 16\,GPa, we observe a clear change in the pressure evolution of the exchange interactions: $J_1$ levels off, and most of the subleading couplings begin to decrease or even turn negative (ferromagnetic). The phase transition from trigonal to monoclinic space group leads to significant changes in the exchange couplings.

\begin{figure*}[t]
\includegraphics[width=\textwidth]{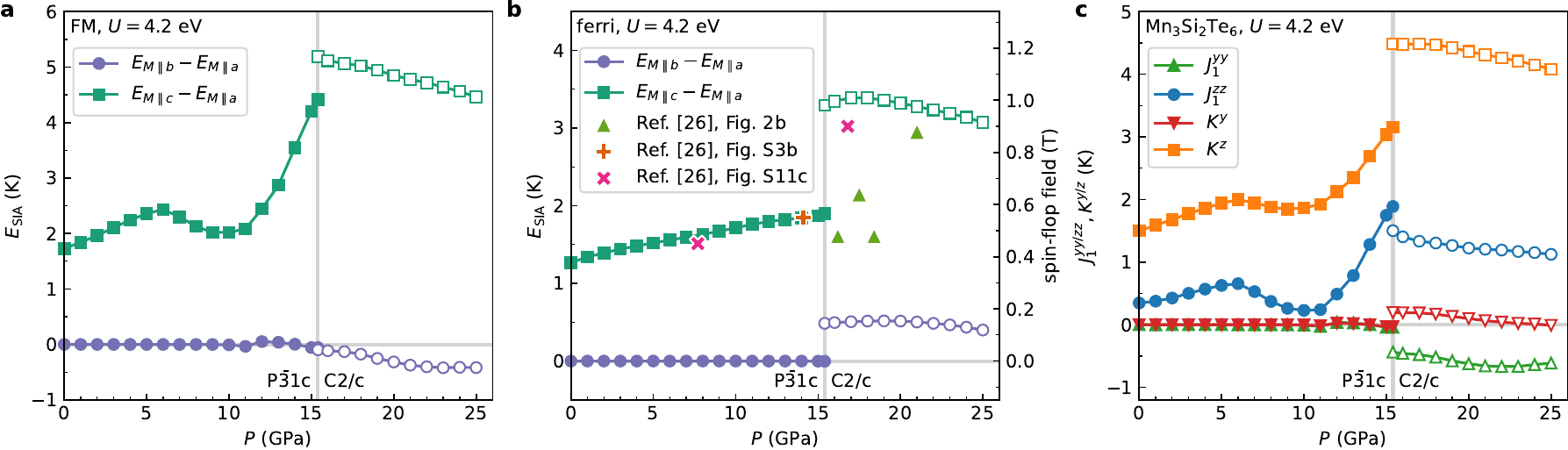}
\caption{\textbf{Single-ion and nearest-neighbour exchange anisotropies of {\mn} from the DFT energy mapping as a function of pressure.} All data points were calculated with $U = 4.2$~eV. Labels $C2/c$ and $P\bar{3}1c$ denote the crystal symmetries on both sides of the phase transition. (a) shows the raw energy differences per Mn atom (also divided by the square of the spin $S=5/2$) for the ferromagnetic state with three possible spin quantization axes and as a function of pressure. (b) shows the raw energy differences per Mn atom (also divided by the square of the spin $S=5/2$) for the ferrimagnetic ground state and compares these energy differences to the experimentally determined $c$ axis spin-flop fields from Ref.~\protect\onlinecite{Susilo2024}. (c) shows the anisotropic single-ion ($K^y$ and $K^z$) and nearest-neighbour exchange ($J_1^{yy}$ and $J_1^{zz}$) parameters as a function of pressure determined from raw energy differences.}
\label{fig:exchange_anisotropies}
\end{figure*}

The dominant coupling $J_1$ that is responsible for the $S=5/2$ trimers remains constant at around 100\,K up to the highest pressure $P=25$\,GPa. The second neighbour coupling $J_2$, which defines a honeycomb network for Mn1, splits into three distinct couplings; while $J_2$ was near zero in the trigonal structure, possibly due to a compensation between ferromagnetic and antiferromagnetic contributions to the exchange, the three couplings $J_2'$, $J_2''$ and $J_2'''$ in the monoclinic structure take on substantial ferromagnetic values. The coupling $J_3$, which defines a buckled honeycomb network connecting Mn1 and Mn2, also splits into three couplings at the transition to monoclinic symmetry. They remain mostly antiferromagnetic but are significantly smaller than $J_3$ before the transition. The average of $J_3'$, $J_3''$ and $J_3'''$ shows an almost linear decrease. The coupling $J_{4a}$ connects second nearest neighbours in the Mn1 honeycomb network while $J_{4b}$ form a Mn2 triangular lattice. Both are small in $P\bar{3}1c$ but become significant and ferromagnetic in $C2c$. The coupling $J_5$ is the second nearest neighbour in the Mn trimer at twice the length of $J_1$. Small and antiferromagnetic in trigonal symmetry (slightly destabilizing the up-down-up order in the trimer), it becomes increasingly ferromagnetic in monoclinic symmetry where it stabilizes the up-down-up trimer order. Across the pressure range, the sub-leading couplings introduce some frustration into the system, which agrees well with the findings in Ref.~\cite{Ye2022}.

To determine the anisotropies from DFT, we calculate the total energy in ferri- and ferromagnetic spin configuration as a function of the spin quantization axis. We then perform an additional energy mapping, considering that the isotropic part of the Hamiltonian $H_\text{iso}$ (Eq.~\ref{eq:isohamil}) is independent of the quantization axis. This allows us to resolve the anisotropy terms in
\begin{equation}
\begin{aligned}
    H = H_\text{iso} 
        &+ \sum\limits_{i < j} \left( J_{ij}^{yy} S_i^y S_j^y + J_{ij}^{zz} S_i^z S_j^z \right) \\
      &+ \sum\limits_{i} \left( K^y (S_i^y)^2 + K^z (S_i^z)^2 \right) \,,
\end{aligned}
\label{eq:anisohamil}
\end{equation}
where $J_{ij}^{yy}$ and $J_{ij}^{zz}$ are nearest neighbour exchange anisotropies and $K_i^y$ and $K_i^z$ are single-ion anisotropies. In the trigonal phase we have $J_{ij}^{yy} = 0$ and $K_i^y = 0$ due to the crystal symmetry.

The DFT energies, evaluated as a function of pressure, magnetic ordering pattern (FM or ferrimagnetic), and spin quantization axis (see Fig.~\ref{fig:exchange_anisotropies}a and b), clearly indicate that ferrimagnetic ordering is more stable than FM, and that the $ab$ plane constitutes the easy plane in trigonal {\mn}. In the monoclinic phase, the energies of the $a$ and $b$ directions split (see Fig.~\ref{fig:exchange_anisotropies}b). Alignment parallel to the $b$ axis yields an energy slightly above the $a$ axis. The energy distance with respect to the $c$ axis increases, so that $c$ clearly remains the hard axis. The ferromagnetic state, which is much higher in energy than the ferrimagnetic state, shows a similar trend (see Fig.~\ref{fig:exchange_anisotropies}a), although the $b$ axis is slightly lower in energy than the $a$ axis. In order to ensure that this finding is not dependent on our choice of interaction strength of $U=4.2$\,eV, we performed a sample calculation at $U=5$\,eV for {\mn} at $P=20$\,GPa. We find exactly the same hierarchy of energies, and the energy differences differ only insignificantly from those shown in Figs.~\ref{fig:exchange_anisotropies}a and b.

The anisotropy energies can be converted into a magnetic field, which is necessary to rotate spins from the easy plane or axis to the hard axis. For this we need to consider the rotation angle between the crystal axes, which differs from $90^\circ$ in the monoclinic phase. These estimates agree very well with the observed spin-flop (coercive) fields in Ref.~\protect\onlinecite{Susilo2024} (see right axis in Fig.~\ref{fig:exchange_anisotropies}b), which is rather puzzling given that Ref.~\protect\onlinecite{Susilo2024} claims a $c$ easy axis in the monoclinic phase. It also seems that some experimental spin-flop fields assigned to pressures above $P_c$ are similar to our calculations in the trigonal phase. This could indicate some uncertainty in the experimentally applied pressures or a sample-dependent critical pressure. We return to the issue of magnetic anisotropy in the discussion section.

As expected from the energies, in the trigonal phase we find positive exchange and single-ion anisotropies for the $z$ direction, while the anisotropies for the $y$ direction are basically zero, except for small convergence errors in the DFT energies (see Fig.~\ref{fig:exchange_anisotropies}c). This agrees well with polarized neutron powder diffraction experiments at ambient pressure~\cite{Baral2025}. Since nearest neighbour spins in the ferrimagnetic trimer of the ground state are antiparallel, the energy difference between $ab$ plane and $c$ axis is mostly due to the single-ion anisotropy $K^z$, while the exchange anisotropy $J^{zz}$ actually decreases the anisotropy energy.

This qualitative picture for the parameters associated with the $c$ axis persists into the monoclinic phase: the exchange anisotropy $J^{zz}$ remains almost constant, while the single-ion anisotropy $K^z$ increases slightly (see Fig.~\ref{fig:exchange_anisotropies}c). The parameters associated with the $b$ axis become non-zero: while the positive single-ion anisotropy $K^y$ is quite small, the anisotropy energy for the $b$ axis is generated mostly by the negative exchange anisotropy $J^{yy}$.

\section{Discussion}
Our results for the electronic structure of {\mn} reproduce the experimentally observed IMT at the structural phase transition. Interestingly, the average charge gap of the magnetic configurations in our study decreases smoothly as a function of pressure and finally vanishes at the structural phase transition (see Fig.~\ref{fig:couplings}c). This result seems to agree with the smooth decrease of longitudinal resistivity as a function of pressure observed in experiment~\cite{Susilo2024}.

Although our calculations faithfully describe the experimentally known electronic and magnetic properties of {\mn} on both sides of the structural phase transition, we cannot fully explain the experimentally observed dome in the anomalous Hall conductivity as a function of pressure (see Fig.~\ref{fig:ahcwithdoping}) upon entering the monoclinic metallic phase~\cite{Susilo2024}. Since the AHE is measured in large magnetic field far above the spin-flop field of $\sim 1$~T (see Fig.~\ref{fig:exchange_anisotropies}b), we can assume the spins to be oriented along the hard axis ($c$ axis). Therefore, the experimental AHC should be compared to Fig.~\ref{fig:ahcwithdoping}b or potentially Fig.~\ref{fig:ahcwithdoping}c. DFT predicts a large intrinsic contribution with negative sign to the xy component of the AHC, while experiment shows small positive values, which decrease between about 16 and 21~GPa~\cite{Susilo2024}. Based on our findings in Fig.~\ref{fig:ahcwithdoping}b it seems possible that the experimentally observed behaviour could be the result of partial cancellation between the large negative intrinsic contribution and a (hypothetical) positive extrinsic contribution of about the same size.  A similar qualitative mismatch has been observed for the closely related van der Waals ferromagnet {\cgt}, where discrepancies have been attributed to extrinsic effects~\cite{Scharf2025, Scharf2025b, Zhang2024b}. Alternatively, the experimentally observed AHC could be the consequence of electron doping, e.g.~due to Te deficiency, as shown in Fig.~\ref{fig:ahcwithdoping}c. A shift of the chemical potential to $\mu = 0.16$~eV produces a dome-shaped AHC remarkably similar to experiment. Eventually, the actual high-pressure structure may differ from the $C2/c$ candidate employed in our calculations. This issue merits further investigation.

\begin{figure}[t]
    \includegraphics[width=\columnwidth]{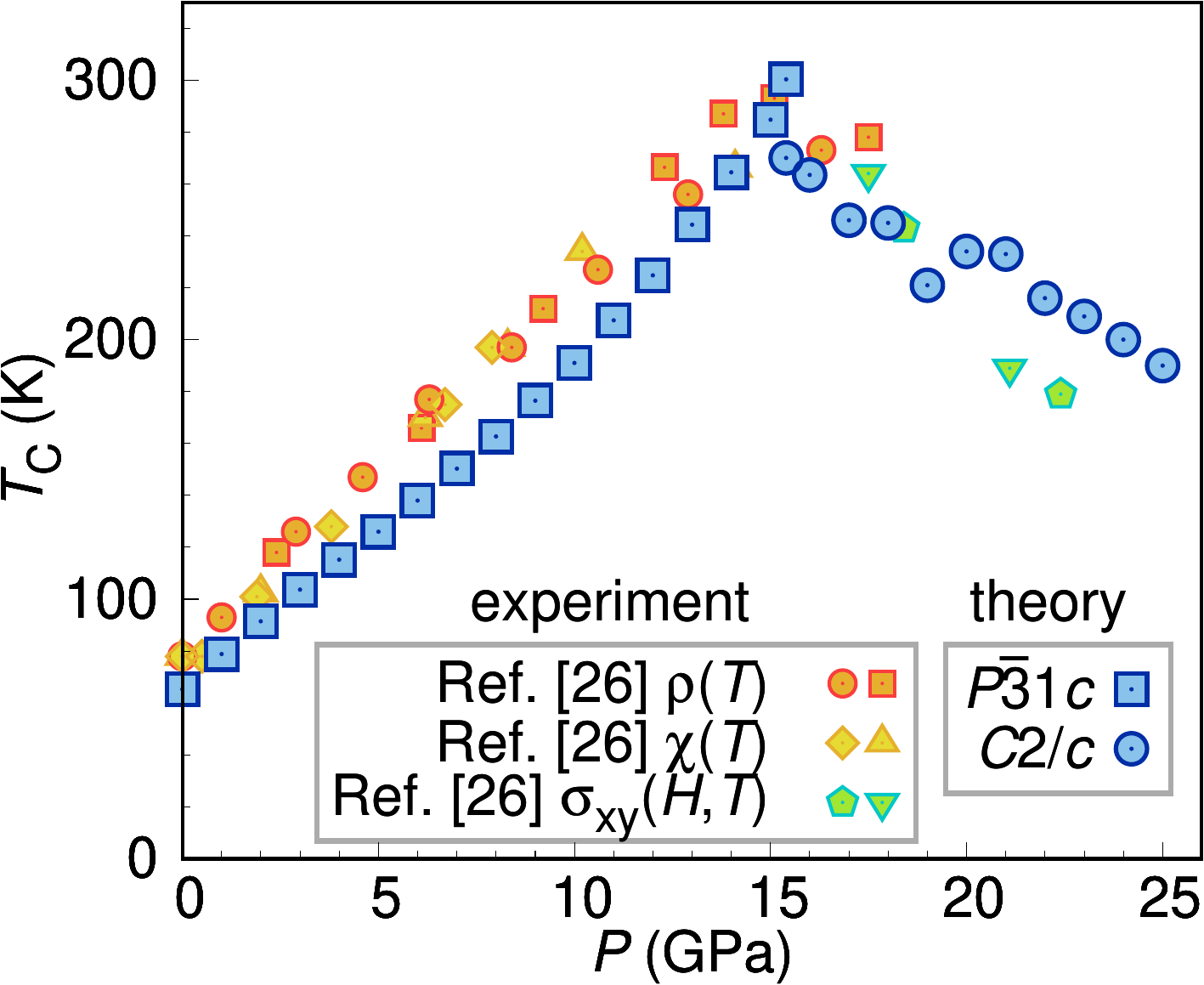}
    \caption{\textbf{Predicted ferrimagnetic ordering temperatures for {\mn} as a function of pressure.} Theoretical values are shown in blue, while experimental data from Ref.~\protect\onlinecite{Susilo2024} are shown in yellow, orange and green symbols for comparison. The theoretical prediction is based on the isotropic Hamiltonian determined from the DFT energy mapping, which we use to simulate magnetic configurations in classical Monte Carlo.} 
    \label{fig:Tc}
\end{figure}

Our results for the isotropic exchange couplings of {\mn} show a rather simple picture of relatively unfrustrated antiferromagnetic exchange in the trigonal high-symmetry structure. In the metallic monoclinic low-symmetry structure, many exchange paths contribute with opposite signs (see Fig.~\ref{fig:couplings}f), which potentially leads to cancellation of contributions and a de-stabilization of the ferrimagnetic ground state as well as a lower ordering temperature. In this sense, the decreasing ordering temperature (see Fig.~\ref{fig:Tc}) cannot be attributed to the decrease or increase of a single exchange coupling. It is rather the product of qualitatively different exchange Hamiltonians in the trigonal vs.~monoclinic structures. Nevertheless, our classical Monte Carlo simulations for the ordering temperature (see Fig.~\ref{fig:Tc}) and the ground state faithfully reproduce the experimentally observed behaviour, also in the low-symmetry phase whose structure is not yet resolved with absolute certainty. This indicates that the $C2/c$ structure is a good candidate for the low-symmetry phase, where the complex cancellation of ferromagnetic and antiferromagnetic exchange paths seems to be the root cause of the decreased ordering temperature.

Our estimates of the exchange interactions and single-ion anisotropies are consistent with multiple studies at ambient pressure~\cite{Bigi2023, Sala2022, Zhang2023}, as well as with experimentally observed spin-flop (coercive) fields under pressure~\cite{Susilo2024}. Our calculations indicate that {\mn} in the low-pressure high-symmetry phase exhibits an $ab$ easy plane and a hard axis along the $c$ direction (see Fig.~\ref{fig:exchange_anisotropies}b). Upon entering the monoclinic phase, the degeneracy between the $a$ and $b$ directions is lifted: the $a$ direction becomes the easy axis, while the $b$ direction lies only slightly higher in energy. The $c$ direction remains the hard axis.

To the best of our knowledge, Ref.~\protect\onlinecite{Susilo2024} reports a $c$ easy axis based on an extrapolation of the magnetic anisotropy energy in the trigonal phase and two low-field ($0.2$~T) magnetization measurements along the $c$ axis. In contrast, our calculations unambiguously identify the $a$ direction as the easy axis in the monoclinic phase and show no indication of a decreasing magnetic anisotropy energy. Consequently, our results are in clear contradiction with the $c$ easy axis proposed in Ref.~\protect\onlinecite{Susilo2024}.

The behaviour of individual anisotropy parameters is rather complex and non-monotonic as a function of pressure (see Fig.~\ref{fig:exchange_anisotropies}c). Nevertheless, the resulting spin-flop fields (see Fig.~\ref{fig:exchange_anisotropies}b) quantitatively agree with experimental data from Ref.~\protect\onlinecite{Susilo2024}. With a spin-flop field of $\sim 1$~T and the strong effect of the orientation of the spin quantization axis on electronic structure, anomalous Hall effect and magnetic properties, comparisons between experiment and theory need to carefully consider the size of magnetic field applied in experiment and its implications for the spin quantization axis in DFT.

In summary, our combined DFT and cMC analysis provides a coherent microscopic picture of the pressure-induced behaviour in {\mn}. The evolution of isotropic and anisotropic exchange couplings across the structural phase transition successfully reproduces the experimentally observed variation of the ferrimagnetic ordering temperature and the spin-flop field. The calculated electronic structure captures the pressure-driven insulator–metal transition, demonstrating an intimate connection between electronic band structure and magnetic properties of {\mn}. However, the experimentally observed dome in the anomalous Hall conductivity cannot be accounted for by intrinsic Berry curvature effects alone, suggesting additional extrinsic contributions or charge doping of experimental samples. {\mn} emerges as a model platform for exploring how pressure couples lattice and spin degrees of freedom with band structure topology in van der Waals magnets.

\section{Methods}
\subsection{Density functional theory-based energy mapping}
We perform density functional theory (DFT) calculations within the full potential local orbital (FPLO) method~\cite{Koepernik1999} and using the generalized gradient approximation (GGA)~\cite{Perdew1996} for the exchange-correlation functional. We account for the strong interactions on Mn$^{2+}$ ions using a DFT+$U$ correction~\cite{Liechtenstein1995}. We use collinear DFT to determine the Heisenberg Hamiltonian parameters, the single ion anisotropy energies and selected anisotropic exchange interactions. The experimental crystal structures under pressure were imported from Ref.~\protect\onlinecite{Susilo2024} and interpolated smoothly using splines (for details see Supplementary Note 1 with Supplementary Figures 1 to 3).

The DFT-based energy mapping approach requires accurate DFT energies for a large set of distinct spin configurations, which are then used to fit the exchange parameters between the spin-5/2 manganese atoms in Eq.~\ref{eq:isohamil} by matching the classical energies of the Heisenberg Hamiltonian with the calculated DFT energies using the method of ordinary least squares (OLS). 

In the trigonal $P\bar{3}1c$ space group, we use $2\times1\times1$ supercells with 12 distinct Mn$^{2+}$ ions. Calculating 30 spin configuruations, we can resolve 14 exchange interactions up to a distance of about three times the nearest neighbour Mn-Mn distance. In the monoclinic $C2c$ space group, we ensure low statistical error bars on the exchange interactions by calculating 100 spin configurations in order to resolve 16 exchange couplings. 

For the monoclinic phase, we observe that parameter estimates for the isotropic Heisenberg couplings can be sensitive to small deviations in the DFT energies. Therefore, we minimize the mean squared error plus a regularization parameter $\alpha$ times an $L_1$ penalty term, which enforces sparsity of the resulting parameter set.~\cite{Tibshirani1996} We find the optimal regularization parameter $\alpha=0.04$ from the onset of the mean squared error as a function of $\alpha$.~\cite{Hastie2009} For more details, see Supplementary Note 3 in Ref.~\cite{SI}, which includes Supplementary Figures 13 to 17.

Single-ion and exchange anisotropies were extracted from DFT by rotating the spin quantization axis and calculating total energies in both ferri- and ferromagnetic spin configuration. Using six DFT energies (from the two spin configurations multiplied by the three choices for the spin quantization axis), we can solve the corresponding system of linear equations for the anisotropies.

In the trigonal phase, we know that the $a$ and $b$ directions are equivalent. Therefore, we have $J_1^{yy} = 0$ and $K^y = 0$. The classical energies of the Hamiltonian with anisotropic terms (Eq.~\ref{eq:anisohamil}) in each spin configuration and direction of spin quantization axis are therefore given by:
\renewcommand{\arraystretch}{1.4} 
\begin{equation}
\begin{pmatrix}
E_{M \parallel a}^\text{FM} \\
E_{M \parallel b}^\text{FM} \\
E_{M \parallel c}^\text{FM} \\
E_{M \parallel a}^\text{ferri} \\
E_{M \parallel b}^\text{ferri} \\
E_{M \parallel c}^\text{ferri}
\end{pmatrix}
=
\begin{pmatrix}
1 & 1 & 0 & 0 \\
1 & 1 & 0 & 0 \\
1 & 1 & n S^{2} & n S^{2} \\
1 & 0 & 0 & 0 \\
1 & 0 & 0 & 0 \\
1 & 0 & -n S^{2} & n S^{2}
\end{pmatrix}
\begin{pmatrix}
E_{0} \\
\Delta_\text{iso}^\text{FM} \\
J_1^{zz}\\
K^z
\end{pmatrix}
\,.
\label{eq:systemanisotrigonal}
\end{equation}
\renewcommand{\arraystretch}{1} 
Due to the crystal symmetry, the energies in $a$ and $b$ direction should be the same. In DFT estimates for the energies on the left-hand side, small deviations may occur due to imperfect convergence.

In the high-pressure monoclinic phase, the $a$ and $b$ directions are not equivalent any more. Therefore, finite anisotropy parameters in the y direction occur. The system of equations in the monoclinic phase reads:
\renewcommand{\arraystretch}{1.4} 
\begin{equation}
\begin{pmatrix}
E_{M \parallel a}^\text{FM} \\
E_{M \parallel b}^\text{FM} \\
E_{M \parallel c}^\text{FM} \\
E_{M \parallel a}^\text{ferri} \\
E_{M \parallel b}^\text{ferri} \\
E_{M \parallel c}^\text{ferri}
\end{pmatrix}
=
\begin{pmatrix}
1 & 1 & 0 & 0 & 0 & 0 \\
1 & 1 & n S^{2} & 0 & n S^{2} & 0 \\
1 & 1 & 0 & n S^{2} & 0 & n S^{2} \\
1 & 0 & 0 & 0 & 0 & 0 \\
1 & 0 & -n S^{2} & 0 & n S^{2} & 0 \\
1 & 0 & 0 & -n S^{2} & 0 & n S^{2}
\end{pmatrix}
\begin{pmatrix}
E_{0} \\
\Delta_\text{iso}^\text{FM} \\
J_{1}^{yy} \\
J_{1}^{zz} \\
K^{y} \\
K^{z}
\end{pmatrix}
\,.
\label{eq:systemanisomonoclinic}
\end{equation}
\renewcommand{\arraystretch}{1} 

We use Eq.~\ref{eq:systemanisomonoclinic} in both the trigonal and monoclinic phase and solve for the unknown parameters on the right-hand side using ordinary least squares. This also allows us to check the quality of our DFT energies, since perfect convergence should provide zero anisotropies in the trigonal phase for $J_1^{yy}$ and $K^y$. Fig.~\ref{fig:exchange_anisotropies}c shows that we recover the zero values for $J_1^{yy}$ and $K^y$ almost perfectly.

From the known DFT energies in the ferrimagnetic state $E_{M \parallel \{ a,b,c\}}^\text{ferri}$, we can also estimate the spin-flop field, i.e.~the magnetic field strength needed to rotate the spins in the direction of the magnetic field. The energy of a spin with $S=5/2$ in a magnetic field is $H_\text{field} = - g \mu_B {\bf B} \cdot {\bf S}$. With ${\bf S} = \frac{5}{2} {\bf u}$, ${\bf B} = B {\bf v}$ and $g = 2$, we can write $H_\text{field} = -5 \mu_B B \, {\bf v} \cdot {\bf u}$. The energy difference between two spin orientations ${\bf u}$ and ${\bf w}$ in a fixed magnetic field is therefore $\Delta E_\text{field} = -5 \mu_B B \, {\bf v} \cdot ({\bf u} - {\bf w})$.

Let us now assume that ${\bf w}$ is the low energy axis in the zero field. That means a spin-flop will occur from the ${\bf w}$ direction to the ${\bf u}$ direction if ${\bf B}$ is parallel to ${\bf u}$ and sufficiently large to overcome the energy difference between the two spin orientations in zero field:
\begin{equation}
\Delta E = E_{M \parallel  {\bf u}}^\text{ferri} - E_{M \parallel  {\bf w}}^\text{ferri} < -5 \mu_B B \, {\bf v} \cdot ({\bf u} - {\bf w})
\label{eq:magneticfieldenergy}
\end{equation}
We now assume ${\bf v} \parallel {\bf u}$, where ${\bf u}$ is the high-energy direction in zero field. With ${\bf v} \cdot {\bf w} = \cos (\beta)$, we can solve for the magnetic field required for the spin-flop:
\begin{equation}
B ({\bf w} \to {\bf u}) > \frac{- \Delta E}{5 \mu_B (1 - \cos (\beta))}
\label{eq:spinflopfield}
\end{equation}

The experimental values in Fig.~\ref{fig:exchange_anisotropies}b are extracted from Fig.~S3b of Ref.~\protect\onlinecite{Susilo2024} by taking the magnetic field necessary for full ferrimagnetic spin polarization along the $c$ axis ($\sim 0.55$~T at 14.1~GPa) and from Fig.~2b and Fig.~S11c of Ref.~\protect\onlinecite{Susilo2024} by taking the field along the $c$ axis necessary for saturation of the anomalous Hall conductivity $\sigma_{xy}$. These are $\sim 0.48$~T at 16.2~GPa, $\sim 0.64$~T at 17.5~GPa, $\sim 0.48$~T at 18.4~GPa, and $\sim 0.88$~T at 21.0~GPa for Fig.~2b of Ref.~\protect\onlinecite{Susilo2024} as well as $\sim 0.45$~T at 7.7~GPa and $\sim 0.9$~T at 16.8~GPa for Fig.~S11c of Ref.~\protect\onlinecite{Susilo2024}.

\subsection{Classical Monte Carlo}
We performed classical Monte Carlo (cMC) simulations using the Metropolis-Hastings algorithm with local updates for {\mn} as a function of pressure. We use a supercell of size $8 \times 8 \times 8$ with 3072 Mn sites in total. For each simulation we perform 5000 lattice sweeps for the warmup. Subsequently, we perform 3000 measurements with 20 lattice sweeps for each. Each cMC run is averaged over the measurements of 512 parallel threads. Therefore, we perform $3072 \cdot (5000 + 3000 \cdot 20) \cdot 512 \approx 10^{11}$ Metropolis steps for each cMC result.

The exchange couplings illustrated in Fig.~\ref{fig:couplings} were used to calculate and plot the specific heat, magnetic susceptibility and ordering temperatures. At ambient pressure, we find a peak in specific heat at 65\,K (see Ref.~\cite{SI}). Upon investigating the equilibrium spin configurations from our cMC data, we constructed polar plots of the spins in the ensemble. This allowed us to confirm that the ground state is ferrimagnetic, and the peak in specific heat corresponds to the ferrimagnetic transition temperature {\tc}~\cite{SI}. We observe a nearly linear increase in {\tc} as a function of pressure in the trigonal phase. In the monoclinic phase, we anticipate a consequent nearly linear decrease in {\tc} based on experimental observations, which is indeed reflected in our Monte Carlo calculations in this phase. Hence, we successfully reproduced the dome-shaped behaviour in {\tc} as a function of pressure through our {\tc} simulations.

\subsection{Anomalous Hall conductivity}
We perform full-relativistic DFT electronic structure calculations, i.e.~including spin-orbit coupling (SOC), based on which we estimate the intrinsic component of the anomalous Hall effect. The anomalous Hall conductivity $\sigma_{xy}$ is defined as the integral of the total Berry curvature $\Omega_{z}(\vec{k})$ over the entire Brillouin zone (BZ)~\cite{Wang2006}:
\begin{equation}
\sigma_{xy} = -\frac{e^2}{\hbar} \int_\text{BZ} \frac{d\vec{k}}{(2 \pi)^3} \, \Omega_z (\vec{k})
\label{eq:conductivityBZintegral}
\end{equation}
The total Berry curvature $\Omega_z (\vec{k})$, calculated using Wannier interpolation within FPLO~\cite{Eschrig2009,Koepernik2023}, is defined as the sum over all bands $n$ of the band-resolved Berry curvature $\Omega_{n, z}(\vec{k})$ weighted by the respective occupation number $f_n(\vec k)$~\cite{Wang2006}:
\begin{equation}
\Omega_z (\vec k) = \sum\limits_n f_n(\vec{k}) \, \Omega_{n, z} (\vec{k})
\label{eq:totalBerrycurvature}
\end{equation}
 
As in our previous study for the similar van der Waals (vdW) ferromagnet CrGeTe$_3$~\cite{Scharf2025}, we calculated the integral over the Brillouin zone in Eq.~\eqref{eq:conductivityBZintegral} using the \texttt{vegas} adaptive Monte Carlo algorithm~\cite{Lepage1978, Lepage2021}. 

For each calculation of the anomalous Hall conductivity $\sigma_{xy}$ we performed ten independent Monte Carlo runs with $10^6$ evaluations of the integrand for training the adaptive part of the algorithm and subsequent $10^6$ evaluations for the actual calculation of the conductivity. The ten independent runs allow us to estimate the standard deviation of the obtained results, i.e.~the Monte Carlo uncertainty.

\section*{Data availability}
The data and scripts for all figures within this study are deposited at Zenodo~\cite{Zenodo} under the following URL: \url{https://doi.org/10.5281/zenodo.18376083}

\section*{Code availability}
The DFT code used in this study is available from \url{https://www.fplo.de}. The
code for the cMC calculations of this study is available from the corresponding author upon request.

\section*{Author contributions}

The project was conceived by DG and HOJ. MS developed the classical Monte Carlo code. VV, MS, DG and HOJ performed the calculations. VV, DG and HOJ analysed and discussed the results and drafted the manuscript. All authors revised the manuscript.

\section*{Competing interests}

The authors declare no competing interests.

\section*{Acknowledgments}

We are grateful to Resta A.~Susilo for help with the experimental crystal structures. H.~O.~J. acknowledges support through JSPS KAKENHI Grants No.~24H01668 and No.~25K08460.  M.~S. is supported by ISHIZUE 2025 of Kyoto University. Part of the computation in this work has been done using the facilities of the Supercomputer Center, the Institute for Solid State Physics, the University of Tokyo. The authors gratefully acknowledge the computing time provided on the supercomputers Lise and Emmy at NHR@ZIB and NHR@Göttingen as part of the German NHR infrastructure.

\bibliography{Mn3Si2Te6}

\clearpage
\includepdf[pages=1]{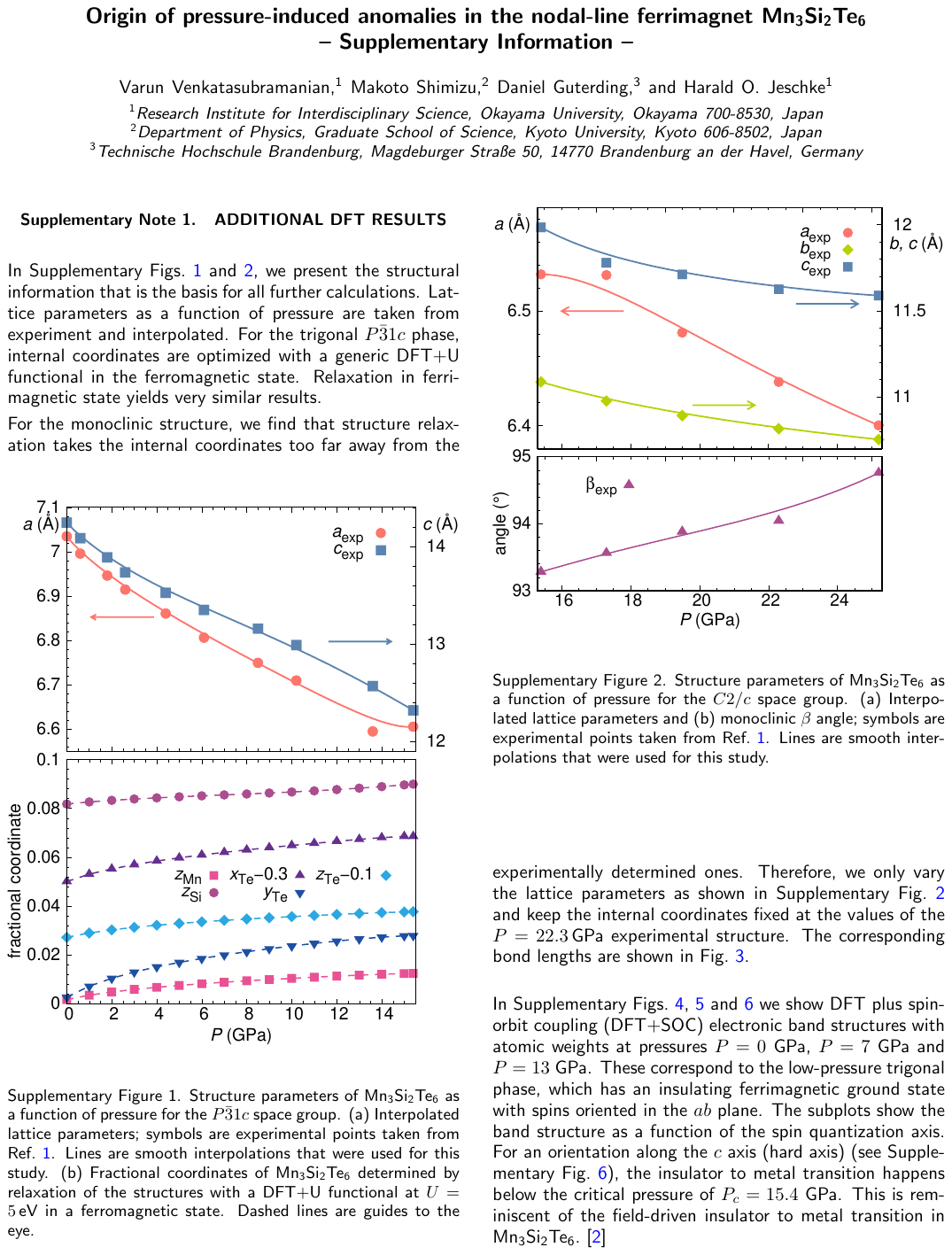}
\clearpage
\includepdf[pages=2]{supplement.pdf}
\clearpage
\includepdf[pages=3]{supplement.pdf}
\clearpage
\includepdf[pages=4]{supplement.pdf}
\clearpage
\includepdf[pages=5]{supplement.pdf}
\clearpage
\includepdf[pages=6]{supplement.pdf}
\clearpage
\includepdf[pages=7]{supplement.pdf}
\clearpage
\includepdf[pages=8]{supplement.pdf}
\clearpage
\includepdf[pages=9]{supplement.pdf}

\end{document}